\documentstyle[12pt,epsfig]{article}
\begin{document}
\begin{titlepage}
\begin{flushleft}
\today
\end{flushleft}
\vskip 5cm
\begin{center}
{\Large\bf Symmetry-Breaking in a $\sigma$-Meson Plasma .\footnote{Supported by Deutsche Forschungs Gemeinschaft under grant number Hu 233/4-3.}}
\vskip 1cm
\mbox{F.M.C.\ Witte}\\
{
\it
Institut f\"ur theoretische Physik\
Universit\"at Heidelberg\\
Philosophenweg  19, \ 69120, \ Heidelberg \\
Germany}
\end{center}
\vskip 1.5cm
\begin{abstract}
In this brief paper we adress spontaneous symmetry breaking in a finite-temperature scalar meson plasma. We calculate the in-medium
averaged thermal $\sigma-\sigma$ scattering crossection and the
related shear viscosity $\eta(T)$ and mean-free-path $L(T)$. 
Our results suggest that slightly below the critical temperature
there is a 30 percent peak in the crossection leading to
equivalent dips in $\eta(T)$ and $L(T)$. We discuss the relevance
of this observation.
\end{abstract}
\end{titlepage}
\newpage

\section{Introduction}
\par
Relativistic Heavy Ion collisions \cite{csernai} most likely offer an
experimental possibillity to study the restoration
of chiral symmetry in hot and highly excited hadronic matter.
Since particularly in the early stage of such a collision the
many-particle system is out of equilibrium it is of interest
to study the effect of a phasetransition on the non-equilibrium
properties of the system.
\par
The simplest model available to study phasetransitions in
a fieldtheoretic setting is the one-component $\varphi^{4}$-model. 
It describes the self-interaction of neutral scalar particles and allows
for the spontaneous breaking of the discrete ${\cal{Z}}_{2}$
symmetry, $\varphi \rightarrow - \varphi$. It has a wide
range of applications. For example, it can be utilized
to study inflationary era's in the early universe \cite{infl}.
However, in the setting of this paper it can be thought of as 
the $\sigma$-meson sector of models like QHD \cite{blat},
as an approximation to the isoscalar boundstates occuring in quark 
models like the NJL-model \cite{jap}, or as the effective color-singlet
field used for example in the Friedberg-Lee model \cite{free} to include
some of the long-range, non-perturbative, non-abelian effects of QCD. 
We will use the $\varphi^{4}$-model and calculate the effect of the phasetransition on the thermally averaged 
crossections, the mean-free-path and the shear-viscosity in a 
$\sigma$-meson plasma.
\par
In this context it is not particularly meaningful
to calculate loop-corrected vertices since there is no reason to
believe that they correctly capture the physics of the mesonic
self-interactions. Rather, the tree-level value of the coupling and
the mass are to be regarded as free parameters of the model.
Loop-corrections to the propagators are neccesary in order
to describe the phasetransition which occurs when the thermal
mass of the mesons vanishes at the critical temperature.
Their effect is most conveniently included by using Hartree-Fock
propagators. Consequently the approach we follow is that we discuss the
contributions of collisions only at lowest-order in the vertices 
while using Hartree-Fock improved propagators.
\par
In this paper we present results that were obtained within a
Closed-Time-Path formulation of the non-equilibrium dynamics
of the $\varphi^{4}$-model \cite{witte}. However the choice
of formalism is arbitrary, and we will therefor not dwell
on it. Extensive treatments can be found in \cite{Ramm}.
In the first section we briefly recapitulate the main steps leading
to the self-consistency equation for the thermal mass of the 
$\sigma$-mesons. Section 2 is devoted to the calculation of the
thermally averaged crossection, the mean-free-path and the shear
viscosity.
 
\section{Spontaneous Symmetry Breaking}
\par
The Lagrangian density of the $\varphi^{4}$-model reads
\begin{equation}
{\cal{L}} = \frac{1}{2} \partial_{\mu}\varphi \partial^{\mu}\varphi
- \frac{m_{B}^2}{2} \varphi^{2} - \frac{\lambda}{4!}\varphi^{4} \ ,
\end{equation} 
and due to $m_{B}^2 < 0$ the ${\cal{Z}}_{2}$ symmetry will be spontaneously broken below the critical temperature and the scalar field will have a non-vanishing expectation value $\varphi_{0}(T)$. At sufficiently high temperatures the symmetry is restorated, i.e. $\varphi_{0}(T \geq T_{c}) = 0$ again. In the Hartree-Fock approximation, 
\begin{equation}
< \varphi^{4} > \approx < \varphi^{2} > < \varphi^{2} > \ ,
\end{equation}
the phasetransition is described by the equations for the expectation
value $\varphi_{0}$
\begin{equation}
\label{fi}
m^2 \varphi_{0} = - \frac{\lambda}{3!} \varphi_{0} \{ \varphi_{0}^2 + 3 G_{0}(x=0) \} \ ,
\end{equation} 
and the Green function $G_{0}$
\begin{equation}
\imath ( p^2 - m^2 ) G_{0}(p) -  \frac{\lambda}{2} \int \frac{d^{4}p '}{(2 \pi )^{4}} \{G_{0}(p ') +  (\varphi_{0})^2\delta^4(p') \} G_{0}(p ) = 1 \ .
\end{equation}
In a thermalized system its solution is just a free thermal Green function with a temperature dependent physical mass $m_{ph}(T)$ mass
\begin{equation}
G_{0}(p ) = \frac{1}{ p^{2} + \frac{1}{4}P^{2} - m_{ph}^{2} } + 2 \pi \imath \delta(p^2 - m_{ph}^2) f(p^{0}) 
\end{equation}
where $f(p^{0})$ is the Bose-Einstein distribution function. The physical mass of the bosons is ofcourse the pole of their propagator. We obtain a consistency equation on $m_{ph}$ by requiring it to be a solution of the Green functions equation. This is the so-called gap equation
\begin{equation}
m_{B}^{2} + \frac{\lambda}{2} \int \frac{d^{4}p '}{(2 \pi )^{4}} \{
G_{0}(p ') +  (\varphi_{0})^2\delta^4(p') \} = m_{ph}^{2} \ ,
\end{equation}
it must be solved along with Eq.(\ref{fi}). The integral of the Green function over momentum is divergent due to the vacuum contribution, and thus the gap equation needs regularization. We calculate the renormalized gap equation via dimensional regularization. The divergence in the integral is absorbed into the bare mass $m_{B}$. We have three masses in the calculation; the bare mass, the physical mass and the renormalized mass. The bare mass is the infinite mass-parameter in the Lagrangian and cannot be measured. The physical mass, $m_{ph}$, is just the pole of the propagator. The renormalized mass, $m_{R}$, is finite and differs from the physical mass by finite terms depending on the renormalization scheme. We use the minimal subtraction scheme and this fixes the finite contributions to $m_{R}$ uniquely. The renormalized mass is in general a function of a scale parameter 
$\mu$. This dependence is fixed by the renormalization group equation for the
running renormalized mass and by a measurement of the mass at some scale. We fix
it by prescribing a zero-temperature mass of $0.5$ GeV. For convenience we introduce the shorthand
\begin{equation}
M_{T}^{2} = \int \frac{d^{4}p '}{(2 \pi )^{4}} 2 \pi \delta((p')^2 - m_{ph}^2)f(p_{0}') \ ,
\end{equation}
 The renormalized gap equation is now found to be
\begin{equation}
m_{R}^{2}(\mu) + \lambda_{R} m_{ph}^{2} \{ \gamma_{E} - 1\} +\frac{\lambda  m_{ph}^{2}}{2} log [ \frac{m_{ph}^{2}}{4 \pi \mu^{2}}] + \frac{\lambda_{R}}{2}(\varphi_{0})^{2} + \frac{\lambda_{R}}{2} M_{T}^{2} = m_{ph}^{2} \ .
\end{equation}
In order to proceed we must first solve the Eq.(\ref{fi}) for 
$\varphi_{0}$. Observe that $\varphi_{0} \neq 0$ implies
\begin{equation}
 m_{B}^2 + \frac{\lambda}{2}G(x=0) < 0 \ .
\end{equation}
This criterium determines the range of temperatures for which the renormalized mass is negative, i.e. the ${\cal{Z}}_{2}$ symmetry is spontaneously broken. The final form of the gap equation reads
\begin{equation}
\label{rengap}
m_{R}^{2}(\mu) + \lambda_{R} m_{ph}^{2} \{ \gamma_{E} - 1\} +\frac{\lambda  m_{ph}^{2}}{2} log [ \frac{m_{ph}^{2}}{4 \pi \mu^{2}}] + \frac{\lambda_{R}}{2} M_{T}^{2} = \left\{ \begin{array}{r} m_{ph}^{2} \ , \ T > T_{c} \\
- \frac{1}{2}m_{ph}^{2} \ , \ T < T_{c} \\ \end{array} \right.
\end{equation}
In the broken phase $\varphi_{0}$ is finite and decreasing as the temperature rises, in the symmetric phase $\varphi_{0}$ vanishes. The physical mass is positive for all temperatures, and vanishes at the critical temperature. 
A typical set of values we used is $\lambda = 5$, $m_{ph}(T=0) = 0.5$GeV which
then yields a critical temperature $T_{c} \approx 3.8$ GeV. With increasing values of $\lambda$ the numerical solution of the gap equation becomes more and more time consuming. Therefore we will resort to a number of scaling arguments to extend our calculation to $\lambda \approx 80$.
\par
With increasing coupling the critical temperature decreases. 
If $\lambda_{1}$ and $\lambda_{2}$ are two values for the coupling constant then a 1-loop estimate suggests that the corresponding critical temperatures roughly satisfy
\begin{equation}
\frac{T_{c1}}{T_{c2}} \approx \frac{m_{R}(1)}{m_{R}(2)}\sqrt{\frac{\lambda_{2}}{\lambda_{1}}} \ ,
\end{equation} 
where the renormalized mass must differ for different couplings. The reason for this is the following: we want to compare the same mesons at different couplings, so $m_{ph}(T=0)$ must be the same for different couplings. 
In general we have
\begin{equation}
\frac{\lambda_{1}}{\lambda_{2}} < 1 \ \rightarrow \ \frac{m_{R}(1)}{m_{R}(2)} > 1 \ .
\end{equation} 
As an estimate based on a 1-loop calculation for $T=0$ we use
\begin{equation}
\frac{m_{R}(1)}{m_{R}(2)} \propto \sqrt{\frac{2+\lambda_{2}}{2+\lambda_{1}}} \ .
\end{equation}
Put together and allowing for stronger non-analytical dependence on
$\lambda$ one finds the following estimate
\begin{equation}
\frac{T_{c1}}{T_{c2}} \approx \frac{m_{R}(1)}{m_{R}(2)}\{\sqrt{\frac{\lambda_{2}}{\lambda_{1}}} +
0.02 [\sqrt{\frac{\lambda_{2}}{\lambda_{1}}}]^3 \}\ ,
\end{equation}
where the numerical constant in front of the second term was obtained 
by fitting this form to the values of $T_{c}$ found by solving E.(\ref{rengap})
between $\lambda = 1$ and $\lambda = 5$. To obtain the critical temperature of chiral symmetry restoration $T_{\chi} \approx 0.15$ GeV we need $\lambda \approx 80$. This value corresponds nicely to what one expects from $SO(4)$ linear sigma model calculations. A phenomenon that is observed in the numerical calculations is that larger couplings give rise to a steeper decrease of $m_{ph}(T)$ as $T \rightarrow T_{c}$. This can be understood as a result of the decrease of the critical temperature with growing $\lambda$.

\section{Collisions}
\par
In this section we will compute the mean free path $L$ and shear viscosity $\eta$ of a plasma of $\sigma$-mesons at finite temperature. In particular we will consider, the effect of spontaneous symmetry breaking. This calculation is of some interest in relation to the non-equilibrium physics of quark-meson plasmas \cite{penf}. The mean free path will give us an indication on wether a quartic self-interaction of the mesons is efficient in preventing the mesons from leaving the plasma. 
\par
We will compute the crossection under the assumption that the c.m. frame of scattering is at rest in the medium. This assumption will lead to differential crossections that only depend on the centre of mass energy $\sqrt{s}$ and the momentum-transfer $\sqrt{-t}$. Furthermore we will assume that the plasma can be considered as a dilute gas. An explicit numerical check on our results yields that this approximation is infact very good.
\par
We will solely rely on the lowest-order diagrams for the reasons
discussed in the introduction. In the symmetric phase we therefor only have
a single diagram, whereas in the broken phase two one-meson exchange diagrams
must also be taken along.
\par
The meson-propagator in our model can be viewed as a pole-approximation to a meson-propagator calculated non-perturbatively in an effective chiral 4-fermion theory. We take the value of the coupling constant $\lambda$ to be a parameter which can in principle also be calculated non-perturbatively from such a microscopic model.
\par
In the symmetric phase the total crossection $\sigma_{0}$ computed in the c.m. frame is given by
\begin{equation}
\sigma_{0}(s) = \frac{\lambda^{2}}{16 \pi s} \ .
\end{equation}
In the broken phase the exchange diagrams also contribute both to the s- and the t-channels. The total crossection in the c.m. frame, this yields 
\begin{eqnarray}
\sigma_{1}(s) & = & \frac{\lambda^2}{16 \pi^2 s} \{ 1 - \frac{\lambda \varphi^{2}}{m^2}[\frac{ 1 - log(\frac{s}{2m^2} - 1)}{\frac{s}{m^2} - 1}] \nonumber \\
& + & \frac{\lambda^2 \varphi^{4}}{m^4}[\frac{log(\frac{s}{2m^2} - 1) - 2}{(\frac{s}{m^2} - 1)^2} + \frac{1}{\frac{s}{m^2} - 1}] \} \ . \nonumber \\
\end{eqnarray}
At threshold the two crossections differ by roughly a factor $3$, reflecting the fact that the number of scattering processes contributing to the crossection in the broken phase has tripled. For higher energies the exchange diagrams become more and more supressed and finally an overall $s^{-1}$-behaviour is found. The dependence of the internal boson-propagator in the exchange diagrams on the distribution function vanishes. This is infact a nice property because in R.P.A. calculations based on the NJL-model \cite{penf} the propagator of the mesons also does not depend on the density of these mesons themselves. In the NJL-model temperature dependence enters the meson propagators through the dynamics of the constituent quarks. In this calculation this is reflected by the temperature dependence of the meson mass.
\par
With the full crossection at our disposal we will now compute the thermally averaged crossection as
\begin{equation}
\sigma_{i}(T) = \frac{\int ds \sqrt{s(s-4m^2)}f^{2}(\frac{\sqrt{s}}{2}) \sigma_{i}(s)}{\int ds \sqrt{s(s-4m^2)}f^{2}(\frac{\sqrt{s}}{2})} \ ,
\end{equation}
where the index $i$ refers to the symmetric phase, $i=0$ and the broken phase $i=1$. The result is given in fig.\ref{termo2}, for $\lambda = 5$ and $m(T=0) = 0.5 GeV$. The crossection is calculated by inserting Hartree-Fock propagators in the internal lines and switching from $\sigma_{0}$ to $\sigma_{1}$ as we decrease the temperature below $T_{c}$. The discontinous jump in the crossection signals the phase transition but its origin lies in the Hartree-Fock approximation for the propagators. We expect that this feature vanishes when higher order corrections are taken into account. It therefore does not serve as a physical signal for the phase transition. The local maximum at $T \approx 3.5$ GeV is caused by the rapid increase of the mass $m_{ph}(T)$ as $T$ falls below $T_{c}$. This is an effect which may survive higher corrections since it occurs well away from the critical temperature and will become stronger at larger values of $\lambda$. Here, for $\lambda = 5$ we notice an!
!
 increase of almost 40 percent 
\par
This averaged crossection enters in the computation of the mean-free-path and the shear viscosity. Using the meson density $n(T)$ given by
\begin{equation}
n(T) = \int \frac{d^3k}{(2 \pi)^3} f(\omega_{k}) \ ,
\end{equation}
the mean-free-path is expressed as \cite{reif}
\begin{equation}
L_{i}^{-1} = \sigma_{i}(T) n(T) \ .
\end{equation}
When using the dressed propagators from the Hartree-Fock calculation we get fig.\ref{mean1}.
The shear viscosity has been computed using the dilute gas approximation
\begin{equation}
\eta_{i} = \frac{16 < p >(T)}{15 \sigma(T)_{i}} \ ,
\end{equation}
where $< p >(T)$ denotes the average momentum calculated from
\begin{equation}
< p >(T) = \int \frac{d^3k}{(2 \pi)^3} k f(\omega_{k}) \ .
\end{equation}
In fig.\ref{shr} we have plotted the result as a function of temperature for the $\lambda = 5$ model with the dressed propagators. 
\par
Both the shear viscosity and the mean-free-path show a local minimum caused by
the local maximum of the crossection just below the critical temperature. 
The fact that viscosity is so large, and the mean-free-path stays very long, is typical for weakly interacting particles \cite{Gro}. We have also calculated the scaled quantity $\eta/T^3$. At the critical temperature $\eta(T=T_{c})/T_{c}^{3} \approx 30$. If we scale this up to the coupling relevant for the chiral phase transition we find $\eta_{\chi}(T=T_{\chi})/T_{\chi}^{3} \approx 60$. If we compare this to the value of $7$ found for the scaled quark shear viscosity in \cite{penf}, it suggests that in a quark-$\sigma$ meson plasma the meson self-interaction yields a contribution to the overall shear viscosity which is approximately an order of a magnitude larger.
\par
These calculations were done in the assumption that a dilute gas approximation is valid, i.e.
\begin{equation}
x_{i} = \frac{n^{-\frac{1}{3}}}{L_{i}} < 1 \ .
\end{equation}
If we check the dilute gas approximation for mesons with a temperature independent mass, and $\lambda = 25 $ we see that it is actually very good in both phases. However by scaling the maximum values of these parameters up to $\lambda_{\chi}$ we  find that in the broken phase the dilute gas approximation may break down. Obviously this will modify the local extrema found at smaller couplings.

\section{Conclusions}

\par 
The question we adressed was how the phase transition from the symmetric, high-temperature, phase to the broken, low-temperature, phase affects the collisions in a meson plasma. At lowest order we compared the shear viscosities and mean free paths in the symmetric and in the broken phase. 
The result is that for temperatures slightly below the critical temperature the in-medium crossections show a local maximum signalling the phasetransition. With Hartree-Fock propagators on the internal lines the crossection displays a discontinous jump at the phase transition. These two distinct features have different origins. The discontinuity in the crossection at the phasetransition is due to the fact that at the critical temperature new scattering processes become available. The use of Hartree-Fock propagators, with their inherent proportionality between the effective lowest-order 3-vertex $\lambda \varphi_{0}(T)$ and the thermal mass $m_{ph}(T)$, causes the jump. It is not a physical signal for the phase transition since it depends strongly on the infrared properties of the underlying diagrams. 
The local maximum in the in-medium crossection slightly below the critical temperature is caused by the rapid increase of the thermal mass $m(T)$ as $T$ falls below $T_{c}$. Here the situation is different in the sense that this feature in our plots does not occur at $T_{c}$ and is thus more reliably extracted from our calculations. Furthermore it will become more prominent at larger values of $\lambda$. Consequently we do expect that the minimum in the mean-free-path and shear viscosity caused by the local maximum of the in-medium crossection is a physically interesting signal of the phase transition.
\par
For future work it may be worthwhile to consider a model including 
quarks in its description.
\newpage

\section*{Acknowledgments}
\par
The author would like to thank J\"org H\"ufner for interesting discussions.



\end{document}